\documentclass[letterpaper, 10 pt, conference]{ieeeconf}
\IEEEoverridecommandlockouts

\usepackage{amsfonts,amsmath,amssymb} 

\DeclareMathOperator{\diag}{diag}         
\def\rbb{\mathbb{R}}
\def\trp{^T}
\def\diag{{\rm diag}}

\def\half{\frac{1}{2}}
\usepackage{psfrag,color}
\usepackage{enumerate,cite,latexsym,graphicx}
\newtheorem{theorem}{Theorem}

\title{\LARGE \bf A Possible Implementation of a Direct Coupling Coherent Quantum Observer}

\author{Ian R.~Petersen and Elanor H. Huntington
\thanks{This work was supported by the
Australian Research Council (ARC) and the Chinese Academy of Sciences President’s International Fellowship Initiative (No. 2015DT006). }%
\thanks{Ian R. Petersen is with the School of  Engineering and Information Technology, 
        University of New South Wales at the Australian Defence Force Academy, Canberra ACT 2600, Australia.
         {\tt\small i.r.petersen@gmail.com} } 
\thanks{Elanor H. Huntington is with the 
College of Engineering and Computer Science, The Australian National University, Canberra, ACT 0200,
Australia. Email: Elanor.Huntington@anu.edu.au.}
}%

\begin{document}

\maketitle
\thispagestyle{empty}
\pagestyle{empty}

\begin{abstract}
This paper considers the problem of implementing a previously proposed  direct coupling quantum observer for a closed linear quantum system. This observer is shown to be able to estimate some but not all of the plant variables in a time averaged sense. The paper proposes a possible experimental implementation of the observer plant system using a non-degenerate parametric amplifier. 
\end{abstract}

\section{Introduction} \label{sec:intro}
A number of papers have recently considered the problem of constructing a coherent quantum observer for a quantum system; see \cite{MJ12a,VP9a,EMPUJ6a}. In the coherent quantum observer problem, a quantum plant is coupled to a quantum observer which is also a quantum system. The quantum observer is constructed to be a physically realizable quantum system  so that the system variables of the quantum observer converge in some suitable sense to the system variables of the quantum plant. The papers \cite{PET14Aa,PET14Ba,PET14Ca,PET14Da}  considered the problem of constructing a direct coupling quantum observer for a given quantum system. 

In the papers  \cite{MJ12a,VP9a,PET14Aa,PET14Ca}, the quantum plant under consideration is a linear quantum system. In recent years, there has been considerable interest in the modeling and feedback control of linear quantum systems; e.g., see \cite{JNP1,NJP1,ShP5}.
Such linear quantum systems commonly arise in the area of quantum optics; e.g., see
\cite{GZ00,BR04}. For such linear quantum system models, an important class of quantum control problems are referred to as coherent
quantum feedback control problems; e.g., see \cite{JNP1,NJP1,MaP3,MAB08,ZJ11,VP4,VP5a,HM12}. In these coherent quantum feedback control problems, both the plant and the controller are quantum systems and the controller is typically to be designed to optimize some performance index. The coherent quantum observer problem can be regarded as a special case of the coherent
quantum feedback control problem in which the objective of the observer is to estimate the system variables of the quantum plant.

\begin{figure}[htbp]
\begin{center}
\includegraphics[width=8cm]{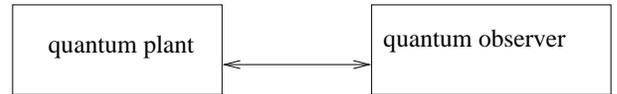}
\end{center}
\caption{Coherent Observer Structure with Direct Coupling.}
\label{F2}
\end{figure}

 In this paper, we consider the situation as in papers \cite{PET14Aa,PET14Ba,PET14Ca,PET14Da} in which there is only direct coupling between quantum plant and the quantum observer. In these papers, both the quantum plant and the quantum observer are assumed to be closed quantum systems which means that they are not subject to quantum noise and are purely deterministic systems. This leads to an observer structure of the form shown in Figure \ref{F2}. In these papers, it is shown that  a quantum observer can be constructed to estimate some but not all of the system variables of the quantum plant. Also, the observer variables converge to the plant variables in a time averaged sense rather than a quantum expectation sense such as considered in the papers \cite{MJ12a,VP9a}.

In this paper, we concentrate on the result presented in \cite{PET14Aa} for the case in which the quantum plant is a single quantum harmonic oscillator and the quantum observer is a single quantum harmonic oscillator. For this case, we show that a possible experimental implementation of the augmented quantum plant and quantum observer system may be constructed using a non-degenerate parametric amplifier (NDPA) which is coupled to a beamsplitter by suitable choice of the amplifier and beamsplitter parameters. 


\section{Quantum Linear Systems}
In this section, we describe the  class of closed linear quantum systems under consideration; see also \cite{JNP1,GJ09,ZJ11,PET14Aa,PET14Ca}. We consider linear non-commutative systems of the form
\begin{eqnarray}
\dot x(t) &=& Ax(t); \quad x(0)=x_0
 \label{linear-c}
\end{eqnarray}
where $A$ is a real matrix in $\rbb^{n
\times n}$, and $ x(t) = [\begin{array}{ccc} x_1(t) & \ldots &
x_n(t)
\end{array}]\trp$ is a vector of self-adjoint possibly
non-commutative system variables; e.g., see \cite{JNP1}. Here $n$ is assumed to be an even number and $\frac{n}{2}$ is the number of modes in the quantum system. 

The initial system variables $x(0)=x_0$ 
are assumed to satisfy the {\em commutation relations}
\begin{equation}
[x_j(0), x_k(0) ] = 2 i \Theta_{jk}, \ \ j,k = 1, \ldots, n,
\label{x-ccr}
\end{equation}
where $\Theta$ is a real antisymmetric matrix with components
$\Theta_{jk}$.  Here, the commutator is defined
by $[A,B]=AB-BA$. In the case of a
single degree of freedom quantum particle, $x=(x_1, x_2)^T$ where
$x_1=q$ is the position operator, and $x_2=p$ is the momentum
operator.  The
commutation relations are  $[q,p]=2 i$.
Here, the matrix $\Theta$ is assumed to be  of the  form
 $\Theta=\diag(J,J,\ldots,J)$ where $J$ denotes the real skew-symmetric $2\times 2$ matrix
$$
J= \left[ \begin{array}{cc} 0 & 1 \\ -1 & 0
\end{array} \right].$$

A linear quantum  system
(\ref{linear-c}) is said to be \emph{physically realizable} if it ensures 
the preservation of the canonical commutation relations (CCRs):
$$x(t)x(t)\trp-(x(t)x(t)\trp)\trp=2i\Theta \ \mbox{for all } t\geq 0.$$
This holds when the system (\ref{linear-c}) corresponds to a collection of \emph{closed quantum
harmonic oscillators}; see \cite{JNP1}.  Such
quantum harmonic oscillators are  described by a quadratic Hamiltonian
$\mathcal{H}=\half x\trp R x$, where $R$ is a real symmetric matrix. 

In the proposed direct coupling coherent quantum observer, the quantum plant is a single quantum harmonic oscillator which is a linear quantum system of the form (\ref{linear-c}) described by the non-commutative differential equation
\begin{eqnarray}
\dot x_p(t) 
&=& A_px_p(t); \quad x_p(0)=x_{0p}; \nonumber \\
z_p(t) &=& C_px_p(t)
 \label{plant}
\end{eqnarray}
where $z_p(t)$ denotes the vector of system variables to be estimated by the observer and $ A_p \in \rbb^{2
\times 2}$, $C_p\in \rbb^{1 \times 2}$. 
It is assumed that this quantum plant corresponds to a plant Hamiltonian
$\mathcal{H}_p=\half x_p\trp R_p x_p$. Here $x_p = \left[\begin{array}{l}q_p\\p_p\end{array}\right]$ where
$q_p$ is the plant position operator and $p_p$ is the plant momentum operator.

We now describe the linear quantum system of the form (\ref{linear-c}) which will correspond to the  quantum observer; see also \cite{JNP1,GJ09,ZJ11}. 
This system is described by a non-commutative differential equation of the form
\begin{eqnarray}
\dot x_o(t) &=& A_ox_o(t);\quad x_o(0)=x_{0o};\nonumber \\
z_o(t) &=& C_ox_o(t)
 \label{observer}
\end{eqnarray}
where the observer output $z_o(t)$ is the  observer estimate and $ A_o \in \rbb^{2
\times 2}$, $C_o\in \rbb^{1 \times 2}$.   Here $x_o = \left[\begin{array}{l}q_o\\p_o\end{array}\right]$ where
$q_o$ is the observer position operator and $p_o$ is the observer momentum operator.  We assume  that the plant variables commute with the observer variables. The system dynamics (\ref{observer}) are determined by the observer system Hamiltonian  
which is a self-adjoint operator on the underlying  Hilbert space for the observer. For the quantum observer under consideration, this Hamiltonian is given by a 
quadratic form:
$\mathcal{H}_o=\half x_o\trp R_o x_o$, where $R_o$ is a real symmetric matrix. Then, the corresponding matrix $A_o$ in 
(\ref{observer}) is given by 
\begin{equation}
A_o=2J R_o \label{eq_coef_cond_Ao}.
\end{equation}
In addition, we define a coupling Hamiltonian which determines the coupling between the quantum plant and the  quantum observer:
\[
\mathcal{H}_c = x_{p}\trp R_{c} x_{o}.
\]

\section{Direct Coupling Distributed Coherent Quantum Observer}

Following \cite{PET14Aa,PET14Ca}, we assume that  $A_p =0$ in (\ref{plant}). This corresponds to $R_p = 0$ in the plant Hamiltonian. It follows from (\ref{plant}) that the plant system variables $x_p(t)$ will remain fixed if the plant is not coupled to the observer. However, when the plant is coupled to the quantum observer this will no longer be the case. In addition, we construct the observer as in \cite{PET14Aa} so that 
\begin{equation}
\label{obs_constants1}
R_o > 0 \mbox{ and }  R_c = C_p^T\beta 
\end{equation}
where
\begin{equation}
\label{obs_constants2}
\beta \in \rbb^{1 \times 2}\mbox{ and } C_oR_o^{-1}\beta^T=-1.
\end{equation}

With this construction of the quantum observer, the following result was established in \cite{PET14Aa}.
\begin{theorem}
\label{T1}
Consider a quantum plant of the form (\ref{plant}) where  $A_p = 0$. Then  the matrices $R_o>0$, $R_c$, $C_o$ satisfying the conditions (\ref{obs_constants1}), (\ref{obs_constants2}) will define direct coupled quantum observer such that for the resulting augmented plant-observer system, the quantity $z_p(t)$ is constant and 
\begin{equation}
\label{average_convergence}
\lim_{T \rightarrow \infty} \frac{1}{T}\int_{0}^{T}(z_p - z_o(t))dt = 0.
\end{equation}
\end{theorem}

\section{A Possible Implementation of the Plant Observer System}
In this section, we describe one possible experimental implementation of the plant-observer system given in the previous section. The plant-observer system is a linear quantum system of the form (\ref{linear-c}) with Hamiltonian 
$\mathcal{H}=\half x\trp R x =  \half x_o\trp R_o x_o + x_{p}\trp R_{c} x_{o}$ where the conditions (\ref{obs_constants1}), (\ref{obs_constants2}) are satisfied. In particular, we assume $R_o = 2\omega_o I > 0$ and hence 
\begin{equation}
\label{total_H}
\mathcal{H} =   \omega_o x_o\trp x_o+ x_{p}\trp C_p^T\beta  x_{o}.
\end{equation} 
Also, the condition (\ref{obs_constants2}) becomes
\begin{equation}
\label{constraint2a}
C_o\beta\trp+2\omega_o =0.
\end{equation}

\begin{figure}[htbp]
\begin{center}
\includegraphics[width=8cm]{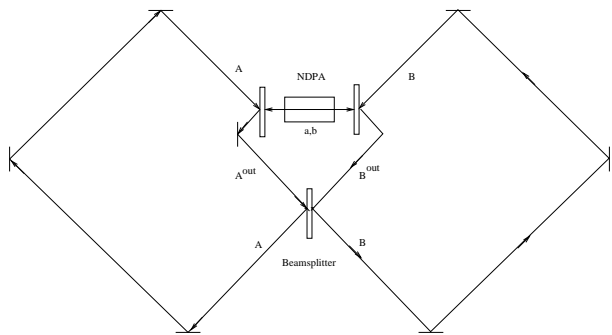}
\end{center}
\caption{NDPA coupled to a beamsplitter.}
\label{F3}
\end{figure}

In order to construct a linear quantum system with a Hamiltonian of this form, we consider an NDPA coupled to a beamsplitter as shown schematically in Figure \ref{F3}; e.g., see \cite{BR04}.

A linearized approximation to the NDPA is defined by a quadratic Hamiltonian of the form
\[
\mathcal{H}_1 = \frac{\imath}{2} \left(\epsilon a^*b^* - \epsilon^* a b\right)+\omega_o b^*b
\]
where $a$ is the annihilation operator corresponding to the first mode of the NDPA and $b$ is the annihilation operator corresponding to the second mode of the NDPA. These modes will be assumed to be of the same frequency but with a different polarization with $a$ corresponding to the quantum plant and $b$ corresponding to the quantum observer. Also, $\epsilon$ is a complex parameter defining the level of squeezing in the NDPA and $\omega_o$ is the detuning frequency of the $b$ mode in the NDPA. The $a$ mode in the NDPA is assumed to be tuned. In addition, the NDPA is defined by the vector of coupling operators $L = \left[\begin{array}{l} \sqrt{\gamma} a \\ \sqrt{\gamma}b \end{array} \right]$. Here $\gamma > 0$ is a scalar parameter determined by the reflectance of the mirrors in the NDPA. 

From the above Hamiltonian and coupling operators, we can calculate the following quantum stochastic differential equations (QSDEs) describing the NDPA:
\begin{eqnarray}
\label{qsde}
\left[\begin{array}{l} d a \\ d b \end{array} \right] &=& 
 \left[\begin{array}{ll} 0 & \frac{\epsilon}{2}  \\ \frac{\epsilon}{2} & 0 \end{array} \right]\left[\begin{array}{l} a^* \\ b^* \end{array} \right]dt \nonumber \\
&&-
 \left[\begin{array}{ll} \frac{\gamma}{2} &  0 \\ 0 & \frac{\gamma}{2}+\imath \omega_o \end{array} \right]\left[\begin{array}{l} a \\ b \end{array} \right]dt \nonumber \\
&&- \left[\begin{array}{ll} \sqrt{\gamma} &  0 \\ 0 & \sqrt{\gamma} \end{array} \right]\left[\begin{array}{l} dA \\ dB \end{array} \right];\nonumber \\
 \left[\begin{array}{l} dA^{out} \\ dB^{out} \end{array} \right] &=& \left[\begin{array}{ll} \sqrt{\gamma} &  0 \\ 0 & \sqrt{\gamma} \end{array} \right]
\left[\begin{array}{l} a \\ b \end{array} \right]dt + \left[\begin{array}{l} dA \\ dB \end{array} \right];
\end{eqnarray}
e.g., see \cite{ShP5}.

 We now consider the equations defining the beamsplitter
 \[
 \left[\begin{array}{l} A \\ B \end{array} \right] = \left[\begin{array}{ll} \cos \theta  & e^{-\imath \phi} \sin \theta  \\ -e^{\imath \phi} \sin \theta  & \cos \theta \end{array} \right]\left[\begin{array}{l} A^{out} \\ B^{out} \end{array} \right]
 \]
 where $\theta$ and $\phi$ are angle parameters defining the beamsplitter; e.g., see \cite{MW95}. This implies
 \[
 \left[\begin{array}{l} A^{out} \\ B^{out} \end{array} \right] = \left[\begin{array}{ll} \cos \theta  & -e^{-\imath \phi} \sin \theta  \\ e^{\imath \phi} \sin \theta  & \cos \theta \end{array} \right] \left[\begin{array}{l} A \\ B \end{array} \right].
 \]
 Substituting this into the second equation in (\ref{qsde}), we obtain
 \begin{eqnarray*}
 \lefteqn{\left[\begin{array}{ll} \cos \theta  & -e^{-\imath \phi} \sin \theta  \\ e^{\imath \phi} \sin \theta  & \cos \theta \end{array} \right] \left[\begin{array}{l} dA \\ dB \end{array} \right] }\nonumber \\
&&= \sqrt{\gamma} \left[\begin{array}{l} a \\ b \end{array} \right]dt + \left[\begin{array}{l} dA \\ dB \end{array} \right]
 \end{eqnarray*}
and hence
\[
\left[\begin{array}{ll} \cos \theta -1   & -e^{-\imath \phi} \sin \theta  \\ e^{\imath \phi} \sin \theta  & \cos  \theta -1 \end{array} \right] \left[\begin{array}{l} dA \\ dB \end{array} \right]=\sqrt{\gamma} \left[\begin{array}{l} a \\ b \end{array} \right]dt.
\]
We now assume that $\cos \theta \neq 1$. It follows that we can write
\begin{eqnarray*}
\lefteqn{\left[\begin{array}{l} dA \\ dB \end{array} \right] =}\nonumber \\
&& \frac{\sqrt{\gamma}}{2(1-\cos \theta)}\left[\begin{array}{ll} \cos \theta -1   & e^{-\imath \phi} \sin \theta  \\ -e^{\imath \phi} \sin \theta  & \cos  \theta -1 \end{array} \right]\left[\begin{array}{l} a \\ b \end{array} \right]dt.
\end{eqnarray*}
Substituting this into the first equation in (\ref{qsde}), we obtain
\begin{eqnarray*}
\label{qsde}
\lefteqn{\left[\begin{array}{l} d a \\ d b \end{array} \right] }\nonumber \\
 &&=\left[\begin{array}{ll} 0 & \frac{\epsilon}{2}  \\ \frac{\epsilon}{2} & 0 \end{array} \right]\left[\begin{array}{l} a^* \\ b^* \end{array} \right]dt \nonumber \\
&&-
 \left[\begin{array}{ll} \frac{\gamma}{2} &  0 \\ 0 & \frac{\gamma}{2}+\imath \omega_o \end{array} \right]\left[\begin{array}{l} a \\ b \end{array} \right]dt \nonumber \\
&&- \frac{\gamma}{2(1-\cos \theta)}\left[\begin{array}{ll} \cos \theta -1   & e^{-\imath \phi} \sin \theta  \\ -e^{\imath \phi} \sin \theta  & \cos  \theta -1 \end{array} \right]\left[\begin{array}{l} a \\ b \end{array} \right]dt.
\end{eqnarray*}
These QSDEs can be written in the form
\[
\left[\begin{array}{l} d a \\ d b \\da^* \\ db^*\end{array} \right] = F \left[\begin{array}{l} a \\ b \\a^* \\ b^*\end{array} \right]dt
\]
where the matrix $F$ is given by 
\[
F = \left[\begin{array}{llll} 
0 & - \frac{\gamma e^{-\imath \phi} \sin \theta}{2(1-\cos \theta)} & 0 & \frac{\epsilon}{2}\\
\frac{\gamma e^{\imath \phi} \sin \theta}{2(1-\cos \theta)} & -\imath \omega_o & \frac{\epsilon}{2} & 0\\
0 & \frac{\epsilon^*}{2} &0 & - \frac{\gamma e^{\imath \phi} \sin \theta}{2(1-\cos \theta)} \\
\frac{\epsilon^*}{2} & 0 & \frac{\gamma e^{-\imath \phi} \sin \theta}{2(1-\cos \theta)} & \imath \omega_o
\end{array} \right].
\]
It now follows from the proof of Theorem 1 in \cite{ShP5} that we can construct a Hamiltonian for this system of the form
\[
\mathcal{H} = \half \left[\begin{array}{llll} a^*& b^*&a & b\end{array} \right]M\left[\begin{array}{l} a \\ b \\a^* \\ b^*\end{array} \right]
\]
where the matrix $M$ is given by 
\[
M = \frac{\imath}{2}\left(JF-F^\dagger J\right)
\]
where $J= \left[\begin{array}{ll}I & 0 \\0 & -I\end{array} \right]$. Then, we calculate
\[
M = \frac{\imath}{2}\left[\begin{array}{llll} 
0 & - \frac{\gamma e^{-\imath \phi} \sin \theta}{1-\cos \theta} & 0 & \epsilon\\
\frac{\gamma e^{\imath \phi} \sin \theta}{1-\cos \theta} & -2\imath \omega_o & \epsilon & 0\\
0 & -\epsilon^* &0 &  \frac{\gamma e^{\imath \phi} \sin \theta}{1-\cos \theta} \\
-\epsilon^* & 0 & -\frac{\gamma e^{-\imath \phi} \sin \theta}{1-\cos \theta} & -2\imath \omega_o
\end{array} \right].
\]

We now wish to calculate the Hamiltonian $\mathcal{H}$ in terms of the quadrature variables defined such that 
\[
\left[\begin{array}{l} a \\ b \\a^* \\ b^*\end{array} \right] = 
\Phi \left[\begin{array}{l} q_p \\ p_p \\q_o \\ p_o\end{array} \right]
\]
where the matrix $\Phi$ is given by
\[
\Phi = \left[\begin{array}{llll} 
1 & \imath & 0 & 0 \\
0 & 0 & 1 & \imath \\
1 & -\imath & 0 & 0 \\
0 & 0 & 1 & -\imath
\end{array} \right].
\]
Then we calculate
\begin{eqnarray*}
\mathcal{H} &=& \half \left[\begin{array}{llll}q_p & p_p &q_o & p_o \end{array} \right]R\left[\begin{array}{l} q_p \\ p_p \\q_o \\ p_o\end{array} \right]\nonumber \\
&=& \half \left[\begin{array}{ll}x_p\trp & x_o\trp \end{array} \right] R\left[\begin{array}{l}x_p \\ x_o \end{array} \right]
\end{eqnarray*}
where the matrix $R$ is given by
\begin{eqnarray*}
R &=& \Phi^\dagger M \Phi\\
&=& \left[\begin{array}{ll} 0 & R_c \\
R_c\trp & 2\omega_o I
\end{array} \right],
\end{eqnarray*}
\[
R_c = \left[\begin{array}{ll}
-\Im(\epsilon) -\Im(\alpha) & \Re(\epsilon)+\Re(\alpha) \\
\Re(\epsilon) -\Re(\alpha) & \Im(\epsilon)-\Im(\alpha)
\end{array} \right]
\]
and $\alpha = \frac{\gamma e^{\imath \phi} \sin \theta}{1-\cos \theta}$. Hence,
\[
\mathcal{H} = \omega_o x_{o}\trp x_{o}+ x_{p}\trp R_{c} x_{o}.
\]
Comparing this with equation (\ref{total_H}), we require that 
\begin{equation}
\label{Rc}
\left[\begin{array}{ll}
-\Im(\epsilon) -\Im(\alpha) & \Re(\epsilon)+\Re(\alpha) \\
\Re(\epsilon) -\Re(\alpha) & \Im(\epsilon)-\Im(\alpha)
\end{array} \right] = C_p^T\beta 
\end{equation}
and the conditions (\ref{obs_constants1}), (\ref{obs_constants2}) to be satisfied in order for the system shown in Figure \ref{F3} to provide an implementation of the augmented plant-observer system. 

\begin{figure}[htbp]
\begin{center}
\includegraphics[width=8cm]{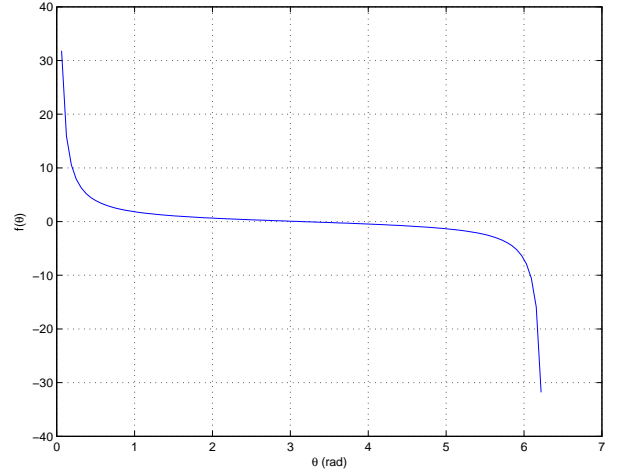}
\end{center}
\caption{Plot of the function $f(\theta)$.}
\label{F4}
\end{figure}

We first observe that the matrix on the right hand side of equation (\ref{Rc}) is a rank one matrix and hence, we require that
\[
\det \left[\begin{array}{ll}
-\Im(\epsilon) -\Im(\alpha) & \Re(\epsilon)+\Re(\alpha) \\
\Re(\epsilon) -\Re(\alpha) & \Im(\epsilon)-\Im(\alpha)
\end{array} \right] = |\alpha|^2 - |\epsilon|^2 = 0.
\]
That is, we require that 
\[
\gamma \left|\frac{\sin \theta}{1-\cos \theta}\right| = |\epsilon|. 
\]
Note that the function $\frac{\sin \theta}{1-\cos \theta}$ takes on all values in $(-\infty,\infty)$ for $\theta \in (0,2\pi)$ and hence, this condition can always be satisfied for a suitable choice of $\theta$. This can be seen in Figure \ref{F4} which shows a plot of the function $f(\theta) = \frac{\sin \theta}{1-\cos \theta}$.

Furthermore, we will assume without loss of generality that $\theta \in (0,\pi)$ and hence we obtain our first design equation
\begin{equation}
\label{theta_eqn}
\frac{\sin \theta}{1-\cos \theta} = \frac{|\epsilon|}{\gamma}.
\end{equation}
In practice, this ratio would be chosen in the range of $\frac{|\epsilon|}{\gamma} \in (0,0.6)$ in order to ensure that the linearized model which is being used is valid. 

Our second design equation is obtained from (\ref{Rc}). First we write $C_p = \left[\begin{array}{ll}
C_{p1} & C_{p2} \end{array}\right]$ and define the complex number $c = C_{p1}+\imath C_{p2}$. The argument of this complex number $\arg(c)$ determines the quadrature of interest in the plant. It is straight forward to verify that the condition (\ref{Rc}) will be satisfied for some non-zero vector $\beta$ if and only if 
\[
\arg\left(\imath(\epsilon -\alpha^*)\right) = \arg(c).
\]
That is,
\[
\arg\left(\epsilon - \frac{\gamma e^{-\imath \phi} \sin \theta}{1-\cos \theta}\right) =\arg(c) - \frac{\pi}{2}.
\]
This is equivalent to 
\begin{equation}
\label{arg1}
\arg\left(\frac{\epsilon}{\gamma} - \frac{ e^{-\imath \phi} \sin \theta}{1-\cos \theta}\right) =\arg(c) - \frac{\pi}{2}.
\end{equation}
Then we write $\epsilon = |\epsilon|e^{\imath \psi}$. It follows from (\ref{theta_eqn}) that (\ref{arg1}) can be re-written as
\begin{equation}
\label{psi_phi_eqn}
\arg\left(e^{\imath \psi}- e^{-\imath \phi}\right) = \arg(c) - \frac{\pi}{2}.
\end{equation}
This is our second design equation. 

If the design equations (\ref{theta_eqn}) and (\ref{psi_phi_eqn})  are satisfied then there will exist a non-zero vector $\beta$ such that (\ref{Rc}) will be satisfied. Then, we can always find a non-zero vector $C_o$ such that (\ref{constraint2a}) is satisfied. Thus, using Theorem \ref{T1} we can conclude that if the the design equations (\ref{theta_eqn}), (\ref{psi_phi_eqn}) are satisfied then the corresponding direct coupled observer will have the desired properties. However, since the proposed experimental implementation of the plant-observer system is a closed quantum system, there is no measurement which can be made on the system to verify the performance of this system. Future research will be directed towards extending the theory developed in \cite{PET14Aa} to allow for open quantum systems in which a measurement can be made to verify the behaviour of the direct coupled quantum observer. 

\noindent
{\bf Example}

We now illustrate the above design principles with an example corresponding to typical laboratory values. In this example, we let $C_p = \left[\begin{array}{ll}1 & 0 \end{array}\right]$ corresponds to the case in which the position quadrature of the plant is of interest. Also, we choose $\gamma = 10^8$ rad/s and $\omega_o = 10^8$ rad/s. In addition, we choose 
\[
\frac{|\epsilon|}{\gamma} = 0.1
\]
which according to (\ref{theta_eqn}) corresponds to a value of $\theta = 168.6^\circ$.

In this case $c=1$ which is purely real and the condition (\ref{psi_phi_eqn}) reduces to 
\[
\arg\left(e^{\imath \psi}- e^{-\imath \phi}\right) = - \frac{\pi}{2}.
\]  
That is
\[
\cos(\psi) -\cos(\phi) = 0
\]
and
\[
\sin(\psi) + \sin(\phi) < 0.
\]
To satisfy these conditions, we choose $\psi = \phi = -90^\circ$. Then, we obtain
\[
\epsilon = 10^7 e^{-\imath\frac{\pi}{2}} = -\imath 10^7
\]
and 
\[
\alpha = 10^7e^{-\imath\frac{\pi}{2}} = -\imath 10^7.
\] 
Then, it follows from (\ref{Rc}) that
\[
\left[\begin{array}{ll}
2 \times 10^7 & 0 \\
0 & 0
\end{array} \right] = \left[\begin{array}{l} 1 \\ 0 \end{array}\right] \beta.
\]
Hence
\[
\beta = \left[\begin{array}{ll} 2\times 10^7 & 0 \end{array}\right].
\]
Therefore, if we choose $C_o = \left[\begin{array}{ll} -10 & 0 \end{array}\right]$, it follows that the condition (\ref{constraint2a}) will be satisfied. Thus, with these parameter values, the proposed implementation will satisfy the conditions of Theorem \ref{T1} for a direct coupled quantum observer. 

\section{Conclusion}
We have shown that the direct coupling quantum observer proposed in \cite{PET14Aa} could be at least in theory implemented experimentally. However, such an experiment could not provide experimental verification that the properties of such a quantum observer described in Theorem \ref{T1} are satisfied. In order to address this issue, future research will extend the results of \cite{PET14Aa} to allow for a small probe field. Then the theory developed in this paper will be extended to allow for this case. 

Another area of possible future research would be to analyse the performance of the proposed implementation of the plant-observer system without making a linearization assumption in the model of the NDPA.


\end{document}